\documentclass[10pt,twocolumn,pra,showpacs,amsmath,
amssymb,aps,floatfix,superscriptaddress,longbibliography]{revtex4-1}

\usepackage{bbm}
\usepackage{empheq}
\usepackage[utf8]{inputenc}
\usepackage{enumerate}

\usepackage{color}

\usepackage{amsthm}
\usepackage{mathrsfs}
\usepackage{textcomp}

\usepackage{graphicx}
\usepackage{epstopdf}

\usepackage[breaklinks=true]{hyperref}
\usepackage{breakcites}

\usepackage{todonotes}

\newcommand{\id}{\mathbbm{1}}                

\begin{document}

\title{Probing quantumness with joint continuous measurements of non-commuting qubit observables}
\author{Luis Pedro Garc\'ia-Pintos}
\affiliation{Institute for Quantum Studies, Chapman University, 1 University Drive, Orange, CA 92866, USA.}
\author{Justin Dressel}
\affiliation{Institute for Quantum Studies, Chapman University, 1 University Drive, Orange, CA 92866, USA.}
\affiliation{Schmid College of Science and Technology, Chapman University, 1 University Drive, Orange, CA 92866, USA.}

\date{\today}

\begin{abstract}
We analyze the continuous measurement of two non-commuting observables for a qubit, and investigate whether the simultaneously observed noisy signals are consistent with the evolution of an equivalent classical system. Following the approach outlined by Leggett and Garg, we show that the readouts violate macrorealistic inequalities for arbitrarily short temporal correlations. Moreover, the derived inequalities are manifestly violated even in the absence of Hamiltonian evolution, unlike for Leggett-Garg inequalities that use a single continuous measurement. Such a violation should indicate the failure of at least one postulate of macrorealism: either physical quantities do not have well defined values at all times, or the measurement process itself disturbs what is being measured. For measurements of equal strength we are able to construct a classical stochastic model for a spin that perfectly emulates both the qubit evolution and the observed noisy signals, thus emulating the violations; interestingly, this model also requires an unphysical noise to emulate the readouts, which effectively restricts the ability of an observer to learn information about the spin. 
\end{abstract}

\maketitle


One of the quintessential features of quantum mechanics is the existence of complementary pairs of observables that do not commute in the operator formalism. Attempting to measure one observable of such a pair necessarily disturbs subsequent measurements made on the other. This sort of unavoidable measurement disturbance departs from the usual classical intuition, since classical observables have definite values that may be probed in principle without observing such disruption.

Leggett and Garg formalized this classical intuition for macroscopic objects by the following two key assumptions of ``macrorealism'' \cite{LeggettGarg85,Leggett1988,Leggett2008}. (i) Macrorealistic systems undergoing causal evolution have observable values that are well-defined at all times. (ii) These values may be measured in principle without disrupting subsequent evolution. From these assumptions it is straightforward to construct inequalities that must be satisfied, which may be broadly construed as analogous to Bell inequalities \cite{Bell,Barbieri2009,Das2013}, but for measurements on a single system that are separated in time rather than space. The noncommutativity of quantum observables prevents these assumptions from being satisfied, so violations of such Leggett-Garg inequalities indicate in some sense the ``quantumness'' of observed system behavior, at least when the measurements are expected to be classically noninvasive on operational grounds \cite{NorireviewLG14,MaroneyLG}. 

Existing protocols that test Leggett-Garg inequalities with quantum systems have encountered noncommutativity in several distinct ways. Variants of the original test proposed by Leggett and Garg \cite{LeggettGarg85,Waldherr2011,Xu2011,Souza2011,Athalye2011,Knee2012,Katiyar2013,BudroniLG15,HuelgaLG15} use a sequence of projective measurements of the same observable, and only show violations if Hamiltonian evolution is placed in between the measurements. Extending this original protocol to time-continuous monitoring of a single observable produces analogous behavior, where concurrent Hamiltonian evolution is still necessary to observe violations~\cite{KorotkovLG,KorotkovLGexperiment,Bednorz2012}. The need for Hamiltonian evolution in these two cases is readily understood in the Heisenberg picture, where evolution is equivalent to passively transforming the measured observable to noncommuting complements over time. Without such evolution, even noisy signals may be interpreted in a purely Bayesian way (i.e., refining the imperfect knowledge of the observer about a definite property over time), and thus will not produce a violation. In contrast, schemes that use a sequence of measurements for different observables \cite{Jordan2006,Williams2008,PrydeLG11,Dressel2011,Fedrizzi2011,Suzuki2012,Groen2013,Dressel2014,White2016} can produce violations without any intermediate Hamiltonian evolution---the measured observables in these protocols are already noncommuting, so there is no way to avoid the intrinsic measurement disturbance.

In this paper we revisit the essence of the Leggett-Garg paradigm and examine the simultaneous time-continuous monitoring of two noncommuting observables ($x$ and $z$) for a qubit \cite{KorotkovXYZ}. This simple example illustrates 
the basic physical issue: attempting to monitor two incompatible properties at the same time produces nontrivial measurement results that seem to violate classical intuition. In contrast to the continuous monitoring of a single observable \cite{KorotkovLG,KorotkovLGexperiment,Bednorz2012}, we find that macrorealistic inequalities are manifestly violated by the combination of the two stochastic measurement records for arbitrarily small correlation intervals; moreover, the violations occur without the need for additional Hamiltonian evolution. When the measurement results are interpreted as noisy records for individual classical spin components, the correlations imply that at all times the spin must be pointing in all directions in the $x$-$z$ plane simultaneously, but also imply that the magnitude of the spin must be zero. These absurd physical conclusions are clearly inconsistent with the assumption that preexisting physical values are being revealed without their measurement being disrupted by some sort of invasiveness. 

In the special case that the two measurements are of equal strength, we are able to resolve the absurdity with an equivalent classical picture that provides a concrete model for the necessary invasiveness. That is, we construct a model for the stochastic evolution of a classical spin that precisely reproduces the absurd macrorealistic conclusions. Perhaps surprisingly, our model involves no nonlinear backaction due to measurement collapse (as in the quantum description), but instead drives the physical spin evolution with intrinsic white noise that can be interpreted as a fluctuating bath related to the detection process. The two seemingly independent measurement records seen by the observer can be constructed from this physical noise, the known evolving state, and an independent white noise that is completely unrelated to the physical evolution. This unphysical white noise effectively enforces an epistemic restriction on what the observer is permitted to know about the physical state evolution \cite{Spekkens2007,Harrigan2010,Bartlett2012}. The stochastic rotations of the qubit in this model occur regardless of whether the observer records the output signals, unlike the usual interpretation of quantum collapse. Though this simple model becomes invalid with asymmetric measurement strengths, it provides an interesting example for how a classical description can appear to be manifestly quantum mechanical.

The paper is organized as follows. In Section~\ref{sec:contmeasurements} we describe the process of continuously monitoring a qubit, first for a single observable ($z$), then for two noncommuting observables ($x$ and $z$). In Section~\ref{sec:LG} we show that the standard assumptions of macrorealism are violated for joint continuous monitoring of both $x$ and $z$, even in the absence of an external drive, and for arbitrarily short correlation times. Nevertheless, in Section~\ref{sec:ClassicalModel} we describe an equivalent stochastic classical system that perfectly reproduces the behavior of the monitored qubit. We discuss our conclusions in Section~\ref{sec:Conclusion}.

\section{Continuous joint measurement of non-commuting observables}
\label{sec:contmeasurements}


Let us start by reviewing the continuous measurement process for a single observable represented by the Hermitian operator $A$. We assume a steady-state detector that consists of a continuous stream of identically prepared Gaussian states \cite{JacobsIntroContMeas06}, each of which briefly interacts with the system for a duration $\delta t$ and is later measured to produce a noisy result $r$. This idea of a continuous measurement is fairly general, but for specificity we will consider a steady-state coherent microwave field in a pumped resonator with rapid decay rate $\kappa$, which very briefly interacts with a transmon qubit for a duration $\delta t \sim \kappa^{-1}$ before escaping the resonator, traveling down a transmission line, being amplified with a phase-sensitive amplifier, and then being measured to produce a homodyne signal \cite{Murch2013,Weber2014,Tan2015,Foroozani2016}. We model this measurement phenomenologically with the quantum Bayesian approach~\cite{KorotkovBayesian1,KorotkovBayesian2,KorotkovBayesian3}, which is equivalent to the optical quantum trajectories formalism~\cite{WisemanBook,diosi1988continuous,Gambetta2008} with coherent resonator states, which disentangle from the qubit in this ``bad cavity'' limit where $\kappa \to \infty$.  

Each segment of the detecting microwave field of duration $\delta t$ interacts with the qubit independently to produce a result $r$, which produces a Markov chain of quantum state updates (equivalent to Bayes' rule) that becomes a stochastic process in the continuum limit. For the sake of simplifying the discussion here, we assume that the collection of the field is perfectly efficient, and that there are no other dephasing or energy-relaxation effects that disrupt the qubit evolution.  We also assume that the measured result $r$ has been scaled such that the probability distribution $P(r|a)$ for obtaining $r$ if the system is in an eigenstate $\left| a \right\rangle$ of $A$ is Gaussian with variance $\tau/\delta t$ centered around the eigenvalue $a$ corresponding to $\left| a \right\rangle$. The measurement time $\tau$ is an experimental parameter that depends on the coupling between the measurement device and the system, and characterizes the rate at which the device acquires information about the state of the system. With this choice of normalization, $\tau$ is the time for an accumulated noisy readout to achieve unit signal-to-noise ratio given a definite initial eigenstate.

Such a Gaussian measurement for observing $r$ during each independent duration $\delta t$ corresponds to a Gaussian positive operator-valued measure (POVM) $E(r)$ that is diagonal in the $A$ basis such that $P(r|a) = \langle a|E(r)|a \rangle$. As such, $E(r)$ satisfies the probability normalization condition $\int E(r)dr = \id$. In absence of experimental inefficiency and phase backaction \cite{KorotkovBayesian2}, each $E(r)$ factors into a single Kraus operator
\begin{equation}\label{eq:gausskraus}
M(r) = \left( \frac{\delta t}{2 \pi \tau }\right)^{1/4} \exp\left[-\frac{ \delta t}{2\tau} \frac{(r - A)^2}{2} \right]
\end{equation}
such that $E(r) = M(r)^\dagger M(r)$. This Kraus operator describes the state update $\rho \mapsto M(r)\rho M(r)^\dagger / \text{Tr}[\rho E(r)]$ resulting from observing a particular $r$, given an initial density matrix $\rho = \sum_{a,a'}\rho_{a,a'}|a\rangle\langle a'|$. 

For simulation purposes, $r$ is a random variable sampled from the mixture distribution $ P(r|\rho(t))=\text{Tr}[\rho(t)E(r)]$ at each time step. In the continuum limit $\delta t \ll \tau$, the readout $r$ approximates a moving-average stochastic process
\begin{align}
\label{eq:readout}
r(t) = \text{Tr}[\rho(t) A] + \sqrt{\tau} \  \xi(t).
\end{align}
That is, the Gaussians with variance $\tau/\delta t$ centered at each eigenvalue $a$ broaden and merge, so the mean of $r(t)$ at each $t$ approximates the mean $\text{Tr}[\rho(t) A]$ of the eigenvalues in the state $\rho(t)$, with the approximately Gaussian spread around that mean becoming additive white noise $\xi(t)$
satisfying $\langle \xi(t) \rangle = 0$ and $\langle \xi(t_1) \xi(t_2) \rangle = \delta(t_1 - t_2) $. Here the averaging $\langle \cdot \rangle$ can denote either a temporal average or an ensemble average since the white noise process is stationary.

Considering two noncommuting observables is a straightforward generalization of the single observable case, obtained by alternating the measurements prior to taking the continuum limit. 
For simplicity we now restrict our discussion to a qubit with Bloch coordinates $x(t) = \text{Tr}[\rho(t)\sigma_x]$, $y(t) = \text{Tr}[\rho(t)\sigma_y]$, and $z(t) = \text{Tr}[\rho(t)\sigma_z]$ defined by the Pauli operators $\sigma_x$, $\sigma_y$, and $\sigma_z$. We will simultaneously measure $x$ and $z$ with equal measurement times $\tau_x = \tau_z = \tau$, in the absence of Hamiltonian evolution. 
The effects of each independent measurement with records $r_x$ and $r_z$ are described by Kraus operators of the form in Eq.~\eqref{eq:gausskraus} with $A = \sigma_x,\,\sigma_z$, which we denote as $M_x(r_x)$ and $M_z(r_z)$, respectively. 
After obtaining both measurements over a timestep $\delta t$, the approximate state update is given by
\begin{equation}
\rho(t+\delta t) \approx  \frac{M_z M_x \rho(t) M_x^\dag M_z^\dag}{\text{Tr}\left[ M_z M_x \rho(t) M_x^\dag M_z^\dag \right]},
\end{equation}
which is valid to first order in $\delta t/\tau \ll 1$. Though this discrete form that performs the two measurements separately will accumulate error of order $(\delta t/\tau)^2$ over time, it is still a useful approximation for numerical simulations, since it properly preserves the properties of the state (unit trace and complete positivity). The accumulated sequencing error may be quantified by comparing the state after the update order ($x$, $z$) to that after the reverse ordering, which will verify whether $\delta t/\tau$ is sufficiently small. In practice, explicitly first order stochastic update methods can accumulate more subtle evolution errors over time without taking proper care of preserving the state properties. 
%

\begin{figure}
    \begin{center}
    \includegraphics[trim={0 1cm 0 0},width=0.5\textwidth]{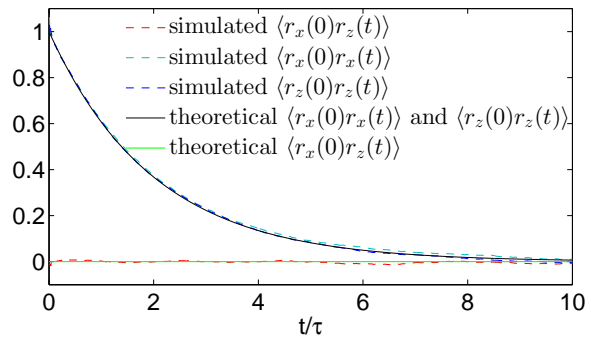}
    \end{center}
\caption{Correlation functions for the independent readout signals $r_x(t)$ and $r_z(t)$ obtained from simultaneously monitoring the noncommuting qubit observables $\sigma_x$ and $\sigma_z$ with equal characteristic measurement times $\tau$. While cross-correlations like $\langle r_x(0) r_z(t) \rangle$ vanish, autocorrelations like $\langle r_x(0) r_x(t) \rangle$ decay exponentially with the delay at rate $1/2\tau$. }
\label{fig:correlators}
\end{figure}

For analytic purposes, expanding the discrete update to linear order and formally taking the continuum limit $\delta t \to 0$ produces a stochastic master equation for $\rho(t)$  
\begin{align}\label{eq:strat}
\dot{\rho} &= \frac{r_x}{\tau}\left[ \frac{\left\{ \sigma_x,\rho \right\}}{2} - x \rho\right] + \frac{r_z}{\tau} \left[\frac{\left\{ \sigma_z,\rho \right\}}{2} - z \rho\right] 
\end{align}
in Stratonovich form (with time-symmetric derivative $\dot{\rho}(t) \equiv \lim_{\delta t\to 0}[\rho(t+\delta t) - \rho(t - \delta t)]/2\delta t$), where we suppress explicit time dependencies for brevity. This form makes it clear that the effect of continuous qubit measurements at each time $t$ is completely described by a renormalized Jordan product $\{A,B\}/2 \equiv (AB + BA)/2$ of each measured observable with the state $\rho$. In Bloch coordinates, this master equation splits into:
\begin{subequations}\label{eq:dynamicsStrato}
\begin{align}
\dot{x} &= \left(1-x^2\right) \frac{r_x}{{\tau}} - x  z  \frac{r_z}{{\tau}}  \label{eq:dynamicsAStrato} \\ 
\dot{y} &= - y  x \frac{r_x}{{\tau}} - y z \frac{r_z}{{\tau}} \label{eq:dynamicsBStrato} \\
\dot{z} &= \left(1-z^2 \right) \frac{r_z}{{\tau}} - x z \frac{r_x}{{\tau}}. \label{eq:dynamicsCStrato}
\end{align}
\end{subequations}
The correlation functions for the observed readouts may be computed from these differential equations~\cite{KorotkovXYZ} 
\begin{align}
\label{eq:correlators}
\langle r_x(0) r_x(t) \rangle &= \langle r_z(0) r_z(t) \rangle = \exp(-t/2\tau) \\ 
\langle r_x(0) r_z(t) \rangle &=  \langle r_z(0) r_x(t) \rangle = 0,  \nonumber 
\end{align}
and match our numerical simulations shown in Fig.~\ref{fig:correlators} for any $t>0$ and any initial qubit state.

It is interesting to compare Eq.~\eqref{eq:strat} with its analogue in It\^o form (with forward-derivative $\dot{\rho} \equiv [\rho(t+\delta t)-\rho(t)]/\delta t$)~\cite{WisemanBook}.
The qubit state evolves in It\^o form according to
\begin{align}\label{eq:Ito}
\dot{\rho} = &-\frac{1}{2\tau}\frac{\big[[\rho,\sigma_x],\sigma_x\big]}{4} -\frac{1}{2\tau}\frac{\big[[\rho,\sigma_z],\sigma_z\big]}{4} \nonumber \\
&+ \frac{\xi_x}{\sqrt{\tau}}\left[ \frac{\left\{ \sigma_x , \rho \right\}}{2} - x \rho \right] + \frac{\xi_z}{\sqrt{\tau}}\left[ \frac{\left\{ \sigma_z , \rho \right\}}{2} - z \rho \right].  
\end{align}
The first two terms are in Lindblad form, and correspond to the ensemble-average dissipation (decoherence) due to the detector, which acts as an external bath on average during the measurement process. The last two terms describe measurement innovation, and are similar to the Stratonovich evolution but involve only the effective white noises $\xi_x$ and $\xi_z$ of each readout (defined as in Eq.~\eqref{eq:readout}; they increase the purity of the state due to the acquisition of information by the measurement devices~\cite{JacobsIntroContMeas06}. For the efficient measurements considered here, the innovation precisely compensates for the dissipation to preserve purity of an initially pure state, which is not apparent in the It\^o picture. However, the relation between individual trajectories and the ensemble average is more clear in the It\^o picture, since the white noise simply averages away. The solutions to Eqs.~\eqref{eq:strat} and \eqref{eq:Ito} are identical, so the choice of derivative definition is a matter of taste.

\begin{figure}[t]
\centering 
\includegraphics[width=0.45\textwidth]{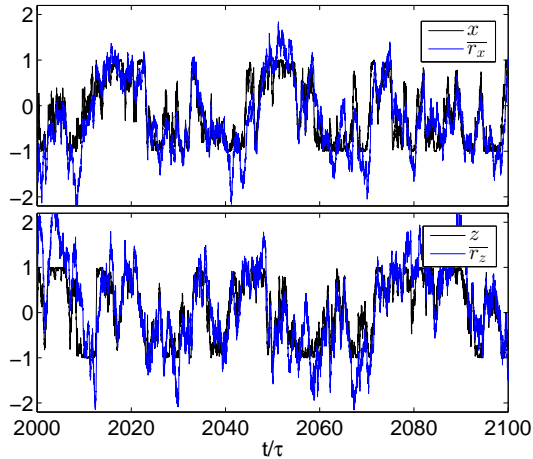} 
\caption{Example filtered output signals $\overline{r}_x(t)$ (top, blue) and $\overline{r}_z(t)$ (bottom, blue), and qubit Bloch coordinates $x(t)$ (top, black) and $z(t)$ (bottom, black), simulated with a normalized time step $\delta t/\tau = 0.01$. The raw readout signals $r_x(t)$ and $r_z(t)$ were filtered with a simple exponentially-weighted moving average with decay time $\tau$, and approximately track the qubit state even without more sophisticated state estimation. 
}
\label{fig:trackedXZ}
\end{figure}

Fig.~\ref{fig:trackedXZ} demonstrates a remarkable feature of the simultaneous measurement of both $x$ and $z$ that is not possible for a single continuous measurement: filtering the raw readout signals $r_x(t)$ and $r_z(t)$ allows the true qubit state $x(t)$ and $z(t)$ to be tracked with reasonably high fidelity \cite{KorotkovXYZ}. For a single measurement this sort of tracking is only possible in the stronger-measurement case where the qubit remains mostly in the eigenstates of the measurement (as in Zeno-pinned quantum jump dynamics \cite{Vijay2011}). Such a single continuous measurement effectively hides the qubit dynamics for timescales shorter than the collapse to the measurement eigenstates (${\sim}3\tau$). However, we see in Fig.~\ref{fig:trackedXZ} that in the two-measurement case even the simplest exponential moving-average filter manages to smooth out the excess readout noise and recover the qualitative qubit state dynamics as $\langle r_x(t)\rangle \approx x(t)$ and $\langle r_z(t) \rangle \approx z(t)$. 

This \emph{model-independent} state estimation method uses the directly observed readouts with minimal processing, and shows that the qubit state no longer collapses to definite eigenstates, but instead seems to behave as if the coordinates $x$ and $z$ are always simultaneously well-defined but also randomly evolving. Importantly, the observer need have no prior knowledge about the qubit to arrive at precisely the same conclusion, with the same estimation of the qubit state. Indeed, Fig.~\ref{fig:trackedXZ} shows an arbitrary evolution segment from a much longer trajectory run, with no further context.

%

This behavior is contrary to what one would naively expect, since we are monitoring two noncommuting observables that should disturb one another. However, we can intuit that the two observables are mutually disrupting the progressive qubit state collapse to their respective eigenstates, such that the disruptions perfectly balance due to the equal $\tau$. This symmetric joint observation thus seems to permit the observables to behave somewhat more \emph{classically}, with both seeming to be reasonably well-defined at all times in an observationally meaningful way.
This behavior is very much in the spirit of the macrorealist assumptions of Leggett and Garg that were discussed in the introduction \cite{LeggettGarg85}, despite the fact that the noncommutativity of the monitored observables is precisely what is expected to be responsible for \emph{causing} violations of such macrorealism.
As such, we are motivated to ask whether the qubit will still violate macrorealistic inequalities for continuous measurements that are similar to existing tests performed with single continuous measurement signals \cite{KorotkovLG,KorotkovLGexperiment,Bednorz2012}.

(As an interesting side note, the Cauchy-Schwartz inequality for qubit observable variances produces
\begin{align}
  (\Delta\sigma_x)^2(\Delta\sigma_z)^2 &\geq \left|\frac{\langle[\sigma_x,\sigma_z]\rangle}{2i}\right|^2 \\
  & + \left|\frac{\langle\{\sigma_x,\sigma_z\}\rangle}{2} - \langle\sigma_x\rangle\langle\sigma_z\rangle\right|^2, \nonumber
\end{align}
which can be rearranged to produce a trivial Bloch sphere inequality $x^2 + y^2 + z^2 \leq 1$. Neither this, nor the coarser Heisenberg-Kennard uncertainty relation derived from it, prevent classical spin-like behavior for a qubit.)

\section{Violation of Leggett-Garg macrorealism}
\label{sec:LG}

In the standard Leggett-Garg scenario one  considers projective measurements of a dichotomic quantity $z(t)$, with $|z(t)|~=~1$, though this restriction can be relaxed to permit $|z(t)|\leq 1$ \cite{Dressel2011,Dressel2014}. Under the assumption that the system obeys \emph{macrorealism}, i.e. that
\begin{enumerate}
\item[(A1)] $z(t)$ evolves causally with a well defined value at any given time $t$ (\emph{macrorealism per-se}), and that
\item[(A2)] $z(t)$ can be measured without disturbing subsequent evolution (\emph{noninvasive measurability}),
\end{enumerate} 
the following three-time inequality holds
\begin{equation}
\label{LGinequalityProj}
\langle z(t_1) z(t_2) \rangle + \langle z(t_2) z(t_3) \rangle - \langle z(t_1) z(t_3) \rangle \le 1, 
\end{equation}
where $\langle \cdot \rangle$ indicates an ensemble average over many realizations of the experiment, each of the realizations consisting of the projective measurement of $z$ at two different times.
Evolving quantum systems can violate the inequality~\eqref{LGinequalityProj}, implying the failure of at least one of the macrorealism postulates.
For qubit measurements of $\sigma_z$ evolving with a Rabi Hamiltonian $H = (\Omega/2) \sigma_x$, the left hand side of~\eqref{LGinequalityProj} becomes $2\cos \left( \Omega \Delta t \right) - \cos \left(\Omega \Delta t\right) = 3/2$ if the time intervals are chosen to be equal such that $t_3 - t_2 = t_2 - t_1 \equiv \Delta t = \pi/2 \Omega$. Note that the violation of the inequality depends crucially on the relation between $\Delta t$ and the period $2\pi/\Omega$---there is no violation in the limit that $\Omega\to 0$ (no evolution).

For continuous monitoring of only $\sigma_z$, this logic is generalized in the following way \cite{KorotkovLG,KorotkovLGexperiment}. First, the noisy measured readout is assumed to be unbiased: $r_z(t) = z(t) + \sqrt{\tau}\,\xi_z(t)$. Second, the noise $\xi_z(t)$ is assumed to be only apparent (i.e., produced by the detector itself) and not causing additional evolution of the qubit (e.g., through measurement backaction, or invasive physical coupling); in this case, $\langle \xi_z(0)z(t) \rangle = 0$ for $t>0$. With this interpretation of continuous \emph{noninvasive measurability}, we can rewrite the correlation functions in Eq.~\eqref{LGinequalityProj} as correlations of the readout directly, $\langle z(t_1)z(t_2)\rangle = \langle r_z(t_1)r_z(t_2)\rangle$, using the fact that the white noise is itself $\delta$-correlated. After this replacement, we recover results completely analogous to the projective measurement case in Eq.~\eqref{LGinequalityProj}.

\begin{figure}[t]
\centering 
\includegraphics[width=0.4\textwidth]{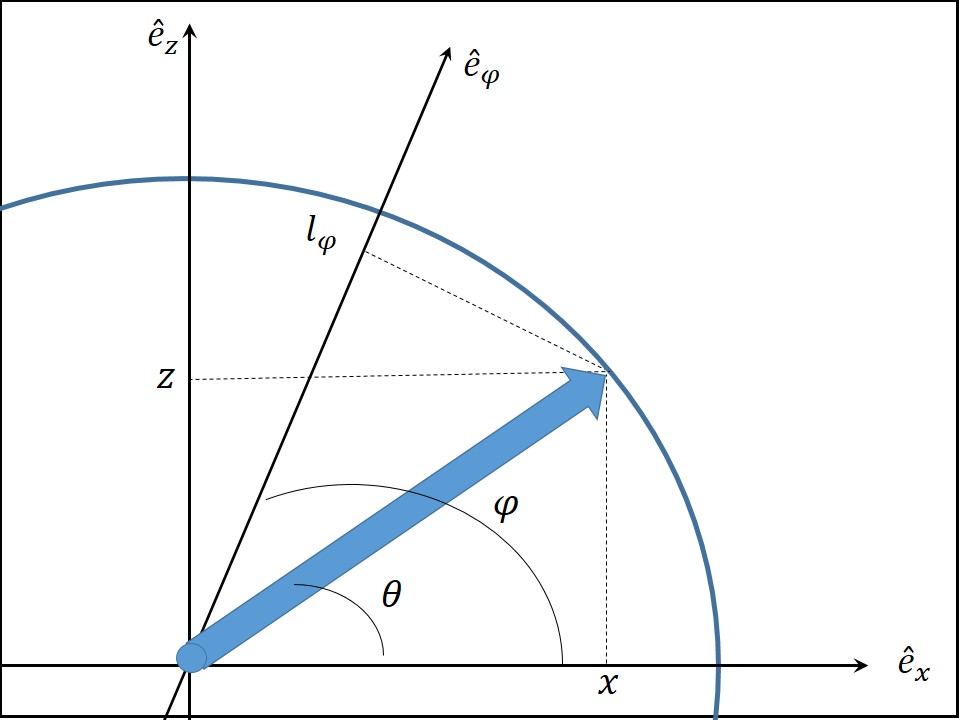} 
\caption{
Bloch representation of a qubit contained in the $y = 0$ plane.
The components $x(t) = \cos\left(\theta(t) \right)$ and $z(t) = \sin\left(\theta(t)\right)$ determine the state at any given time. For a classical spin, these components are sufficient to deduce the component $l_\varphi(t) = x(t)\cos\varphi + z(t)\sin\varphi$ along any direction defined by the angle $\varphi$, as shown.}
\label{fig:spinbloch}
\end{figure}

Let us now assume macrorealism holds for the system being considered, and use the same logic as above to derive suitable macrorealistic constraints for joint $x$ and $z$ monitoring. Since now two orthogonal axes are involved, we expect the macrorealistic state of the qubit to mimic that of a classical spin.
From Fig.~\ref{fig:spinbloch} it is easy to see that from the observed components $x(t)$ and $z(t)$ we can deduce the component $l_\varphi(t)$ of such a definite spin state in an arbitrary direction defined by the angle $\varphi$, 
\begin{equation}
\label{eq:length}
l_\varphi = \cos(\varphi) x + \sin(\varphi) z.
\end{equation}

Similarly, for a given direction $\varphi$ we can construct an effective readout signal for $l_\varphi(t)$ as
\begin{equation}
\label{eq:rphi}
r_{\varphi} = \cos(\varphi) r_x + \sin(\varphi) r_z \equiv l_\varphi + \sqrt{\tau} \ \xi_\varphi,
\end{equation}
where $\xi_\varphi = \cos(\varphi)\xi_x + \sin(\varphi)\xi_z$ is still zero-mean $\delta$-correlated white noise.

If $r_x(t)$ and $r_z(t)$ do convey information about the instantaneous values of $x(t)$ and $z(t)$, then $r_\varphi(t)$ should also provide information about the instantaneous value of $l_\varphi(t)$. It then follows that if we assume \emph{noninvasive measurability} as before, so the apparent noise does not disturb the measured quantity, $\langle \xi_\varphi(0) l_\varphi(t) \rangle = 0$, 
then 
\begin{align}
\label{eq:phiphi}
\langle r_\varphi(0) r_\varphi(t) \rangle &= \langle l_\varphi(0) l_\varphi(t) \rangle.
\end{align}
This is the natural generalization of the continuous Leggett-Garg assumptions to the case of joint continuous measurements. 

From the correlation functions in Eq.~\eqref{eq:correlators} 
the observations for the quantum mechanical model then become
\begin{align}
\label{eq:phiphi1}
\langle r_\varphi(0) r_\varphi(t) \rangle &= 
 \cos^2(\varphi) \langle r_x(0) r_x(t) \rangle + \sin^2(\varphi) \langle r_z(0) r_z(t) \rangle \nonumber  \\
 & + \sin(\varphi) \cos(\varphi) \Big\langle r_x(0) r_z(t) + r_z(0) r_x(t) \Big\rangle  \nonumber \\
 & 
 = \exp(-t/2\tau), \qquad \forall \varphi. 
\end{align}

For short times $t \ll 2\tau$ this implies $\langle r_\varphi(0) r_\varphi(t) \rangle \approx 1$. Since $|l_\varphi| \le 1 $, Eq.~\eqref{eq:phiphi} can only be fulfilled if $|l_\varphi(t)| \approx 1$ for all times and for \emph{any direction $\varphi$}. This is clearly inconsistent with the qubit acting like a spin with a well determined state in the Bloch representation, even before invoking an inequality like Eq.~\eqref{LGinequalityProj}. 

\begin{figure}
    \centering
    \includegraphics[width=0.46\textwidth]{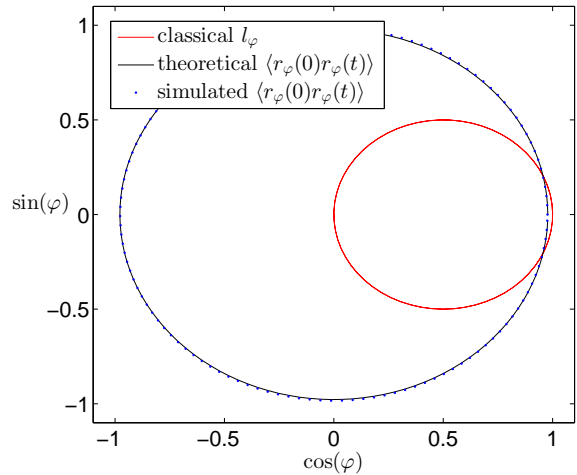}
            \caption{
Comparison of the component $l_\varphi$ for a classical spin pointed along the positive $x$ axis (red), and the correlation results for a qubit, both theoretical (black), and numerically simulated (blue, dotted). 
The quantum correlation functions $\langle r_\varphi(0)  r_\varphi(t) \rangle$ for short delay times $t \ll \tau$ nearly saturate the maximum value of $1$ that $l_\varphi(0)  l_\varphi(t)$ can take at any given time. With the Leggett-Garg noninvasive measurability assumption, this implies that \emph{for all times} and \emph{for all angles} $l_\varphi \approx 1$. This is clearly incompatible with the possible values that the component $l_\varphi$ can take for a classical spin with a definite direction in the Bloch sphere. 
}
\label{fig:quantumvsclassical}
\end{figure}

The incongruities do not end there. Now consider the product of components of the spin along orthogonal directions defined by the angles $\varphi$ and $\varphi + \frac{\pi}{2}$. We obtain, by similar calculations as above, that
\begin{align}
\label{eq:phiphi2}
&\langle r_\varphi(0) r_{\varphi+\pi/2}(t) \rangle = 0 \qquad \forall \ \varphi.
\end{align}
We thus conclude $|l_\varphi(t)| = 0$ for all $t$ and $\phi$, which is incompatible with the previous conclusion that $|l_\varphi(t)| \approx 1$. 
In Fig.~\ref{fig:quantumvsclassical} we check these results with numerical simulations, and compare them to the what one would obtain for a well-defined classical spin.

Notice that with these results it is also simple to construct Leggett-Garg inequalities for non-commuting measurements that are similar to Eq.~\eqref{LGinequalityProj}. 
As an example, invoking the same noninvasive measurability assumption as before we obtain
\begin{align}
\label{LGnewclassical}
\langle r_x(0) r_\varphi(t) \rangle + \langle r_\varphi(t) r_z(2t) \rangle - \langle r_x(0) r_z(2t) \rangle &~\le~1
\end{align}
for any $\varphi$. However, the actual evolution under joint measurement of $x$ and $z$ yields
\begin{align}
&\langle r_x(0) r_\varphi(t) \rangle + \langle r_\varphi(t) r_z(2t) \rangle - 
\langle r_x(0) r_z(2t) \rangle \nonumber \\
&\quad =  \left( \cos(\varphi) + \sin(\varphi)  \right)  \exp(-t/2\tau) \le \sqrt{2}, \label{eq:LGnew}
\end{align}
from using conditions~\eqref{eq:correlators}. For $\varphi = \pi/4$ the right hand side of Eq.~\eqref{eq:LGnew} is approximately $\sqrt{2}$ for $t \ll \tau$, violating the macrorealistic bound~\eqref{LGnewclassical}. Notably, these bound violations occur without Hamiltonian evolution, unlike  Leggett-Garg inequalities formed from a single output, like $r_x$ or $r_z$ independently; as such, in order to see these violations it is crucial to consider both measurement outputs simultaneously.

Following the usual logic for Leggett-Garg inequalities, we infer from these absurd conclusions that at least one of the assumptions of macrorealism is being violated. One option is to reject realism, but this is unlikely given the strongly realistic behavior suggested by Fig.~\eqref{fig:trackedXZ}. It is thus more likely that the standard assumption of noninvasive measurability being used for continuous measurements is overly restrictive. It is thus interesting to compute what the quantum dynamics actually imply about the necessary form of the noise invasiveness in order to reproduce the apparent contradictions above. 

Let us revisit Eqs.~\eqref{eq:phiphi1} and~\eqref{eq:phiphi2} and expand them properly in terms of the quantum model. Since $\xi_x$ and $\xi_z$ are independent white noises, we have $\langle \xi_\varphi(0) \xi_\varphi(t) \rangle = \delta(t)$, and since the values of prior state components do not influence later white noise we also get $\langle l_\varphi(0)~\xi_\varphi(t)\rangle = 0$. As such, the proper correlation expansion that includes the invasiveness of the noise for $t > 0$ is
\begin{align}
\langle r_\varphi(0) r_\varphi(t) \rangle = \langle l_\varphi(0) l_\varphi(t) \rangle + \sqrt{\tau} \langle \xi_\varphi(0) l_\varphi(t)\rangle.
\end{align}
Imposing that the quantum predictions from Eqs.~\eqref{eq:correlators} be satisfied then places the following constraints on the noises. 
%
\begin{subequations}
\begin{align}
\big\langle x(0) x(t) + \sqrt{\tau}\xi_x(0) x(t) \big\rangle &= \exp(-t/2\tau) \label{eq:constraint1} \\
\big\langle z(0) z(t) + \sqrt{\tau}\xi_z(0) z(t) \big\rangle &= \exp(-t/2\tau) \label{eq:constraint2} \\
\big\langle x(0) z(t) + \sqrt{\tau}\xi_x(0) z(t) \big\rangle &= 0 \label{eq:constraint3} \\
\big\langle z(0) x(t) + \sqrt{\tau}\xi_z(0) x(t) \big\rangle &= 0, \label{eq:constraint4}
\end{align}
\end{subequations}
which intertwine the dynamics of the system with the noise output from each measurement device. 
These equations need to be satisfied by any macrorealistic model of the underlying evolution of the quantum state that models the invasiveness of the noise. 

\section{Violating macrorealism with an epistemically restricted classical model}
\label{sec:ClassicalModel}

We will now construct a classical model for a spin in a fluctuating magnetic field that accounts for the effect of the noise, and that perfectly emulates both the dynamics of the qubit and the readout signals output in an experiment. The form of this model is sufficient only for measurements of symmetric strength (equal $\tau$), but it provides interesting insight into the structure of the preceding Leggett-Garg violations.

To derive a classical model, we write the equations of motion for the angle
$\theta(t)$ in the $x$-$z$ plane \cite{KorotkovXYZ,Areeya2013,Areeya2015}, defined for a pure state by $x(t) = \cos(\theta(t))$ and $z(t) = \sin(\theta(t))$. From the Stratonovich Eqs.~\eqref{eq:dynamicsStrato} this angle has the equivalent dynamics 
\begin{align}
\dot{\theta} & = x \dot{z} - z \dot{x} = \frac{ x r_z}{{\tau}} - \frac{z r_x}{{\tau}} \equiv \frac{\widetilde{r}}{\tau},
\end{align}
where we have redefined the noise as
\begin{equation}
\label{eq:rtilde}
\widetilde{r} \equiv x r_z- z r_x.
\end{equation}
Surprisingly, this new noise $\widetilde{r}(t)$ behaves precisely as state-independent white noise, 
\begin{align}
\label{eq:rtildewhite}
&\big\langle \widetilde{r}(0) \widetilde{r}(t) \big\rangle= \sqrt{\tau}\, \delta(t),
\end{align}
which can be shown by noticing $\widetilde{r} = \sqrt{\tau} \big( x \xi_z - z\xi_x\big) $ from the expressions in Eq.~\eqref{eq:readout} for $r_x$ and $r_z$, along with the pure state condition $x^2~+~z^2~=~1$, and that $\xi_x$ and $\xi_z$ are independent white noises. This identity completely eliminates the nonlinear state dependence in the evolution, so that the angular velocity $\dot{\theta}$ instantaneously responds to an arbitrary white noise drive. We can think of a classical magnetic moment $\vec{\mu}$ for a spin in the $x$-$z$ plane with evolution $\dot{\vec{\mu}} \propto \vec{B}(t)\times\vec{\mu}$ due to an environmental magnetic field $\vec{B}(t) = B(t)\hat{y} \propto \widetilde{r}(t) \hat{y}$, with fluctuating magnitude and fixed direction along the $y$-axis. This evolution produces random spin rotations in a similar manner to the random velocity kicks received during the Brownian motion of a particle. Note, however, that for Eq.~\ref{eq:rtildewhite} to produce truly white noise it is crucial that the timescale $\tau$ is the same for both measurements.

\begin{figure}[t]
\centering 
\includegraphics[width=0.4\textwidth]{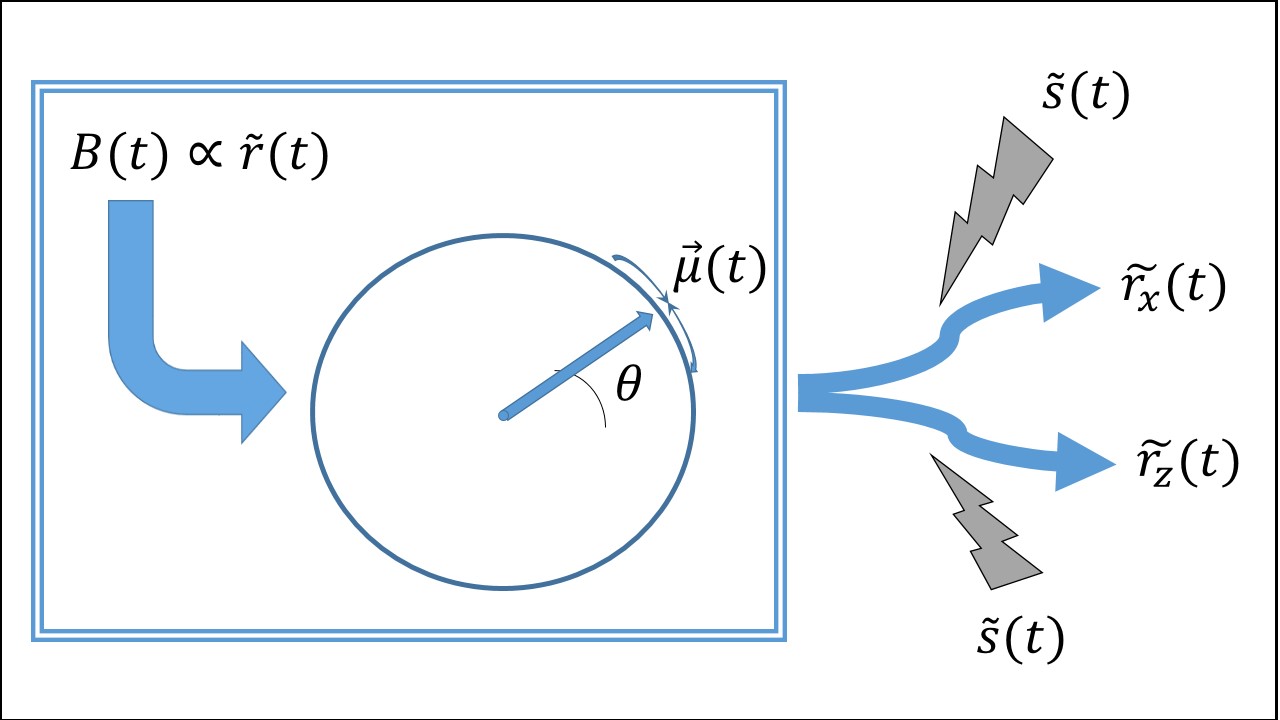}
\caption{Illustration of a classical system that emulates the dynamics of the qubit subjected to the joint continuous measurement of $\sigma_x$ and $\sigma_z$. 
A magnetic field $B(t) \propto \widetilde{r}(t)$, with $\widetilde{r}(t)$ stochastic white noise, drives the magnetic moment $\vec{\mu}$ of the classical spin.
An agent can then combine the driving \emph{physical white noise} $\widetilde{r}$, along with the state of the spin, with an independent \emph{subjective white noise} $\widetilde{s}$ to produce the effective readouts $\widetilde{r}_x  =  x + (-z \widetilde{r} + x\widetilde{s})$ and $\widetilde{r}_z  = z + (x  \widetilde{r} + z\widetilde{s})$, which can be later given to a third party. These effective readouts, as well as the dynamics of the classical spin, perfectly emulate the readout and dynamics expected for a monitored qubit as in Fig.~\ref{fig:trackedXZ}.}
\label{fig:EpistemicModel}
\end{figure}

Now that the spin dynamics have been physically fixed in an observer-independent way by environmental white noise, suppose an agent can measure both the spin angle $\theta(t)$ and the environmental noise $\widetilde{r}(t)$ without disturbing them (i.e., assume true macrorealism). This agent can now construct \emph{effective} readouts $\widetilde{r}_x$ and $\widetilde{r}_z$ from the measured \emph{physical white noise} $\widetilde{r}$ and a second auxiliary \emph{subjective white noise} $\widetilde{s}$ (also satisfying $\langle\widetilde{s}(0)\widetilde{s}(t)\rangle = \sqrt{\tau}\,\delta(t)$) that is known only to the agent. The construction of the effective readouts has the structure of a rotation that inverts the transformation of Eq.~\eqref{eq:rtilde} by mixing the physical and subjective noises 
\begin{align}\label{eq:effectivereadout}
\widetilde{r}_x  =  x + (-z \widetilde{r} + x\widetilde{s}) , \qquad 
\widetilde{r}_z  = z + (x  \widetilde{r} + z\widetilde{s}).
\end{align}
It is then easy to check that these effective readouts satisfy the expected correlation functions, and that averaging the effective readouts will approximately track the state components $x$ and $z$ with additive white noise precisely as illustrated by Fig.~\ref{fig:trackedXZ}. Notice that this is true in spite of the fact that these effective white noises have been constructed from both the physical white noise $\widetilde{r}$ and an unrelated subjective noise $\widetilde{s}$ introduced by the agent. Suppose now that, as depicted on Fig.~\ref{fig:EpistemicModel}, the agent sends these effective readouts to a third party, and informs the third party that they are true measurement records for a continuous qubit measurement. This third party, hampered by the lack of knowledge about the signal preparation, will be unable to find any discrepancies with this claim. As far as the third party will be able to tell, the two readouts $\widetilde{r}_x$ and $\widetilde{r}_z$ will appear to have been generated by the continuous measurement of a qubit. Indeed, the evolution Eqs.~\eqref{eq:dynamicsStrato} can be used by the third party to integrate these readouts and perfectly emulate what will seem like genuine qubit evolution; only the agent will know that these reconstructed ``qubit trajectories'' are actually equivalent to an observer-independent physical spin evolution. 

By construction, the dynamics of the classical spin with these \emph{epistemically restricted effective readouts} will be indistinguishable from those of a qubit undergoing joint continuous measurements of $x$ and $z$. Although the agent has perfect knowledge of the classical spin dynamics, the physical noise, and the irrelevant subjective noise, the third party only receives restricted knowledge that hides the structure of the noise, so would make the same macrorealistic conclusions about the dynamics that were derived in the previous section. This equivalence is consistent with other observations in the literature that quantum models can share many features with epistemically restricted classical models \cite{Spekkens2007,Harrigan2010,Bartlett2012}. 

We emphasize that when the measurement is asymmetric ($\tau_x \neq \tau_z$), then this simple spin model becomes invalid and more complicated classical dynamics will be needed to explain the basins of attraction that appear around the dominant measurement poles (e.g., an additional electric field). Nevertheless, the simplicity of the present model suggests a way to understand how the measured output in Fig.~\ref{fig:trackedXZ} could be consistent with realistic behavior.

\section{Conclusion}
\label{sec:Conclusion}

By considering simultaneous monitoring of both the $x$ and $z$ Bloch coordinates of a qubit, we have shown that
the measured readouts contain structure that challenges the usual application of the notion of Leggett-Garg macrorealism to continuous quantum measurement. Assuming noninvasive measurability---by treating the observed unbiased noise as only apparent and not driving the physical dynamics---the collected readouts manifestly violate macrorealistic inequalities for arbitrarily short correlation times. Interpreted as a spin, such correlations would imply the striking conclusion that the spin points in all directions simultaneously with magnitude one at all times, while also having a magnitude of zero.

To be logically consistent according to macrorealism, one has to admit the possibility that either i) the measurement process is invasive,
with the observed noise having a physical effect on the system, or ii) the physical quantities being measured do not have definite values at all times. Since the qualitative qubit dynamics may be recovered from model-independent averaging of the collected readouts directly, rejecting the latter assumption seems unwarranted in this case. Instead, intrinsic measurement invasiveness seems much more likely.

The apparent invasiveness of the measurement process leaves an imprint, in the form of correlations created between the intrinsic noises from the measurement devices and the physical values being measured. 
Any postulated underlying dynamics for the system are thus constrained by the structure of the correlation functions predicted by quantum mechanics from the collapse postulate. Consistency with quantum predictions is not sufficient to guarantee ``quantumness'' of the mechanism for invasiveness, however. 

To emphasize this point, we constructed an equivalent classical model for a spin undergoing the same dynamics as the qubit, which is valid for the special case of equal measurement strengths for $x$ and $z$. The stochastic evolution is driven by a fluctuating environmental magnetic field, and  produces experimental output that perfectly emulates the records one would obtain from continuously monitoring both $x$ and $z$ coordinates of a qubit. Hence, the output of this classical emulation also violates Leggett-Garg inequalities, and thus seems to violate macrorealism, even though the state of the classical system is well defined and in principle knowable at all times. Importantly, the actual effect of the measurement is not invasive at the level of the observer, since the dynamics of the classical spin and the physical environmental noise are independent from the generation and collection of the observed records. 

To reproduce the qubit measurement records using the classical model, an agent (possibly the measuring device itself) must transform the physical noise driving the evolution to include additional \emph{subjective} noise that has no relevance to the evolution. This extra noise thus constitutes an epistemic restriction on what the observer is allowed to learn about the physical state of the system. That is, the experimental readouts give disguised, as opposed to full, information about the ontic state of the classical system and its physical noise.






\acknowledgments
We thank Alexander Korotkov, Juan Atalaya, Leigh Martin, Shay Hacohen-Gourgy, Irfan Siddiqi and Andrew Jordan for helpful discussions.
This work was supported by US Army Research Office Grant No. W911NF-15-1-0496.
We also acknowledge partial support by Perimeter Institute for Theoretical Physics. Research at Perimeter Institute is supported by the Government of Canada through Industry Canada and by the Province of Ontario through the Ministry of Economic Development \& Innovation.

\bibliography{referencesLGxz}
\end{document}